\documentstyle[dina4,12pt]{article}

\begin{document}
\title{Compatibility of Correlation Dynamics of SU(N) Gauge Theories and
Gauss Law in Temporal Gauge
\thanks{This work was supported in part
by the National Natural Science Foundation and the Doctoral Education
Fund of the State Education Commission of China,
by the Deutsche Forschungsgemeinschaft, GSI Darmstadt, and BMFT.}}

\author{Shun-Jin Wang$^{1,2}$, Wolfgang Cassing$^1$ and Markus H. Thoma$^1$ \\
{\it 1.Institut f\"ur Theoretische Physik, Universit\"at Giessen,}
\\{\it 35392 Giessen, Germany }  \\
{\it 2.Department of Modern Physics, Lanzhou University}\\
{\it Lanzhou 730000, PR China }

\maketitle
\begin{abstract}
A constraint correlation dynamics up to 4-point Green's functions is proposed
for SU(N) gauge theories which reduces the N-body quantum field problem to
the two-body level. The resulting set of nonlinear coupled equations fulfills
all conservation laws including baryon number, linear and angular momenta
as well as the total energy. Apart from the conservation laws in the space-time
degrees of freedom the Gauss law is conserved identically for all times.
The constraint
dynamical equations are highly non-perturbative and thus applicable also in
the strong coupling regime as e.g., low-energy QCD problems.
\end{abstract}
\newpage

A lot of problems in QCD at zero temperature, finite temperature, and in
the case of non-equilibrium situations require the use of non-perturbative
methods. Important examples are the mechanism of confinement, probably
related to the vacuum
stucture, the investigation of hadron spectra and hadronic matter, the
understanding of the nature of the deconfinement transition and the chiral
symmetry restoration. Furthermore, the mechanism of hadronization
in relativistic heavy-ion
collisions following the possible preformation of a quark-gluon plasma as
well as non-perturbative effects in a quark-gluon plasma above the critical
temperature cannot be treated perturbatively as well \cite{r1a}.
Besides numerically
very complicated lattice calculations \cite{r1} there are only preliminary
attempts to incorporate non-perturbative effects by using the Dyson-Schwinger
equation or variational calculations \cite{r2,r2a,r2b}.

In particular, considering the formation, evolution,
and decay of a quark-gluon plasma
in relativistic heavy-ion collisions, a transport theory based on gauge
covariant Wigner functions has been proposed by Elze et al.
\cite{r3a,r3}. It is, in
fact, a many-body theory completely equivalent to the Heisenberg equations
for the quarks and gluon fields, however, as a reformulation of the SU(N)
gauge theory in phase space not managable in practice. The standard
applicable limit is the semi-classical, abelian approximation which
corresponds to lowest order perturbation theory in the high temperature
approximation. A more promising approach may be provided by the use of
equal-time Wigner functions \cite{r4}.

In this letter, we propose a different approach along the line
of relativistic two-body correlation dynamics \cite{r5}-\cite{r9}, which
has proven to provide the genuine basis for the formulation of non-perturbative
transport theories for baryons and mesons \cite{r8}. It is already
known that this approach obeys all conservation laws concerning space-time
degrees of freedom, i.e. fermion number, linear and angular momentum as well
as the total energy \cite{r5}. The novel phenomena in the application to
quarks and gluons is the non-abelian dynamics of the gluon fields as well as
a SU(N) internal gauge symmetry, which we will treat by algebraic methods.

The basic idea of correlation dynamics is the use of dynamical equations of
motion for equal time Green's functions. For this purpose, we consider the
operators
$\hat{G}_n^{(a, \alpha)} (x_1, \cdots , x_n)$, which are products of
the gluon-field operator $A^a_\mu$ and its canonical conjugated momentum
$\Pi^a_\mu$ as well as quark field operators $\Psi^\dagger$ and $\Psi$,
\begin{equation}
\hat{G}_n^{(a,\alpha)}(x_1. \cdots, x_n) = G_n^{(a, \alpha)}(A^a_\mu,
\Pi^a_\mu, \Psi^\dagger, \Psi),
\end{equation}
where $a$ is the color index, $\alpha$ the spinor index, and $n$ the
number of field operators. The equal time Green's functions are given by the
quantum expectation value within the physical state $|p>$
\begin{equation}
G^{(a, \alpha)}_n(x_1, \cdots, x_n) =
<p| \hat{G}_n^{(a, \alpha)} |p> .
\end{equation}

The equations of motion for these Green's functions follow from the
Heisenberg equations
\begin{equation}
\frac{d}{dt} G^{(a, \alpha)}_n(x_1, \cdots, x_n) =
 \frac{1}{i} <p| [\hat{G}_n^{(a, \alpha)}, H] |p> .
\end{equation}
Using canonical quantization rules i.e., equal time commutation relations
between the fields, (3) leads to a coupled set of equations
relating various n-point functions. In order to solve those equations
this set has to be truncated.

A truncation scheme which has been quite successful in the nuclear physics
context is based on the cluster expansion of disconnected Green's functions
in terms of connected (correlated) Green's functions since connected
Green's functions become of minor importance with increasing order \cite{r5}.
Here, we will restrict ourselves to at most 4-point functions, which will
turn out to be the optimum (minimum) truncation scheme for QCD. The
corresponding equations are non-linear and of non-perturbative nature.
They can be easily transformed into transport equations for equal time
Wigner functions by means of a Wigner transformation.

Since the correlation dynamics is based on the Hamilton formulation, it
is inevitable to fix the gauge in the case of SU(N) gauge theories. A well
defined Hamiltonian can be found in temporal gauge \cite{r10}. However,
due to a residual gauge freedom, not all equations of motion result
from the Heisenberg equation, but the Gauss law
\begin{equation}
g^a(x) = J^a(x) + \Psi^\dagger(x) T^a \Psi(x) = 0
\end{equation}
with
\begin{equation}
J^a(x) = \frac{1}{g} D^{a c}_j \Pi^c_j
\end{equation}
is missing. Here we adopt the notation of Christ and Lee \cite{r10}.
In quantum physics (4) cannot be fulfilled as an operator
equation but has to be imposed as a constraint on the physical states
\begin{equation}
g^a(x)\> |p> = 0.
\end{equation}
This equation is equivalent to Slavnov-Taylor-Ward identities, which
account for relations between Green's functions \cite{r11}.

The aim of the present letter is to show the compatibility of the correlation
dynamics equations with the Gauss law and the Ward identities, respectively.
For this purpose, we first note
that the operators $g^a(x)$
at equal times constitute a set of local generators for the SU(N) gauge
group because of \cite{r10}
\begin{equation}
[g^a({\bf r}; t), g^b({\bf r'}, t)] = i f^{a b c} g^c({\bf r}, t)
\ \delta^3({\bf r - r'}) .
\end{equation}
Furthermore, the Gauss law operator commutes with the Hamiltonian \cite{r10}
\begin{equation}
[g^a({\bf r}, t), H] = 0
\end{equation}
which expresses the conservation of $g^a$ in time
\begin{equation}
\frac{d}{dt} g^a(x) = \frac{1}{i} [g^a(x), H] = 0.
\end{equation}
Equations (7) and (8) imply that within the temporal gauge the system
evolved in time by $H$ has a residual gauge symmetry and the
Gauss operators $g^a(x)$ are the local generators of the residual gauge
symmetry.

Now, we assume the
actual realization of the Gauss law as a quantum expectation
value
\begin{equation}
<p|g^a(x)|p> = 0 .
\end{equation}
instead of (6). This may be considered as a weak form of the Gauss law,
since it is less restrictive for the physical states than (6). For example,
the perturbative vacuum $|0>$ fulfills (10) but not (6) \cite{r12}. Also
(10) allows for the local propagation of colored objects, whereas (6)
restricts to color-singlet objects. These consequences of (10) are exactly
what we aim at for describing properties of the quark-gluon plasma.

An actual realization of eq. (10) can be achieved by going over to the
Cartan form of the local algebra $\{g^a(x)\}$, i.e. $\{h^i(x), e^{\alpha}\}
= \{x^b(x)\}$, by means of a nonsingular linear transformation
\begin{equation}
x^b(x) = T^{b a} g^a(x),
\end{equation}
where $\{h^i(x)\}$ are the Cartan operators and $\{e^{\alpha}(x)\}$ are the
raising/lowering operators. If $|p>$ is a common eigenstate of the cartan
operators $\{h^i(x)\}$ with zero eigenvalue,
\begin{equation}
h^i(x) |p> = 0,
\end{equation}
then we immediately obtain from group theory
\begin{equation}
<p|h^i(x)|p> = 0 = <p|e^{\alpha}(x)|p> .
\end{equation}
Since the transformation $T^{a b}$ is nonsingular, we directly get (10).
Thus the weak Gauss law in the general quantum case implies that the physical
state should be a common eigenstate of the Cartan operator $\{h^i(x)\}$
with eigenvalue zero. It should be noted, that although $<p|h^i(x)^2|p>=0$,
$<p|g^a(x)^2|p>$ does not vanish in general, thus allowing for fluctuations.

Furthermore, since all $g^a(x)$ are conserved
quantities and the field equations of motion are generated by $H$, the
weak Gauss law is conserved identically in time provided that it is fulfilled
initially. This is just what we have been aiming at: The Hamiltonian
dynamics are compatible with the conservation of the weak Gauss law, and the
problem is shifted to an initial constraint problem.

Whereas the weak Gauss law is a constraint imposed on the lowest order Green's
functions of the fields by the residual gauge symmetry, the Ward identities
impose constraints on higher order Green's functions due to the same residual
gauge symmetry. We thus will treat the Ward identities also algebraically.

We start defining a Lie operation $L_{g^a(x)}$ by
\begin{equation}
L_{g^a(x)} g^b(x') =: [g^a(x), g^b(x')].
\end{equation}
{}From the algebraic structure of $g^a(x)$ (7) we get
\begin{equation}
L_{g^{a_1}({\bf r_1}, t)} L_{g^{a_2}({\bf r_2},t)} g^{a_3}({\bf r_3}, t) =
(i)^2 \ f^{a_1 b c} f^{a_2 a_3 b} \delta^3({\bf r_2 - r_3})
\delta^3({\bf r_1 - r_2}) g^c({\bf r_1}, t),
\end{equation}
or
\begin{equation}
L_{g^{a_1}({\bf r_1}, t)} L_{g^{a_2}({\bf r_2},t)}
\cdots L_{g^{a_{n-1}}({\bf r_{n-1}}, t)}  g^{a_n}({\bf r_n}, t) =
F^{a_1 \cdots a_n a_{n+1}}({\bf r_1, \cdots, r_n})
 g^{a_{n+1}}({\bf r_1}, t),
\end{equation}
where $F^{a_1 \cdots a_n a_{n+1}}({\bf r_1, \cdots, r_n})$ is a function of
$\{{\bf r_1, \cdots, r_n}\}$ independent of $g^a({\bf r}, t)$.
Thus we obtain from (10)
\begin{equation}
<p|L_{g^{a_1}({\bf r_1}, t)} L_{g^{a_2}({\bf r_2},t)}
 g^{a_3}({\bf r_3}, t)|p> = 0,
\end{equation}
and
\begin{equation}
<p|L_{g^{a_1}({\bf r_1}, t)} L_{g^{a_2}({\bf r_2},t)}
\cdots L_{g^{a_{n-1}}({\bf r_{n-1}}, t)}  g^{a_n}({\bf r_n}, t)|p> = 0 .
\end{equation}

The relation between (17), (18) and gauge invariance can be established
as follows. Since $H$ due to the temporal gauge has a residual gauge
symmetry generated by $\{g^a(x)\}$, the residual group $U(g)$ will generate
a set ${\cal H}_{g p}$ of physical states $|p>_g$ from any physical state
$|p>$
\begin{equation}
|p>_g = U(g) |p>; \  for \ |p>_g \epsilon \ {\cal H}_{p g} ,
\end{equation}
where the residual group element is given by
\begin{equation}
U(g) =  \exp\left (i \int d^3r\> g^b \omega^b \right ),
\end{equation}
where $\omega^b=\omega^b({\bf r})$ are the local gauge transformation
angles.
The state $|p>_g$ has the same energy as $|p>$ since
\begin{equation}
_g<p|H|p>_g = <p|U^{-1}(g) H U(g)|p> = <p| H |p> .
\end{equation}
The weak Gauss law is fulfilled as well
\begin{eqnarray}
_g<p| g^a |p>_g & = & <p|  \exp(-i \int g^b \omega^b \ d^3r)\>
g^a\> \exp(i \int g^c \omega^c \ d^3r) |p> =  \nonumber\\
 <p| U_{a b}\> g^b |p> & = & 0,
\end{eqnarray}
where $U_{a b}$ is a color matrix containing the gauge transformation angles.

{}From the theory of Lie groups it is evident that (18) and (22) are
equivalent. Furthermore, since $U(g)$ generates a gauge transformation for
the SU(N) gauge fields, any gauge invariant physical observable $O$ has
the same expectation value within the set ${\cal H}_{p g}$.

It is interesting to note that the weak Gauss law (10), in contrast to (6),
breaks the linear superposition
principle in ${\cal H}_{p g}$. This comes about as follows: We consider
an arbitrary linear combination of the physical states
\begin{equation}
|\hat{\Psi}> = \int d\mu(g)\> f(g)\> U(g)\, |p>
\end{equation}
with $|\hat{\Psi}>$ normalized to unity and $\mu (g)$ denoting the Haar
measure. Though the Hamiltonian $H$ has the
same energy within the states $|\hat{\Psi}>$,
\begin{equation}
<\hat{\Psi}| H |\hat{\Psi}> = <p| H |p> = E ,
\end{equation}
the weak Gauss law is not fulfilled in general i.e,
\begin{equation}
<\hat{\Psi}| g^a(x) |\hat{\Psi}> = \int d\mu(g') d\mu(g)
<p| U^{-1}(g') g^a(x) U(g) |p> f^*(g') f(g)  \neq 0
\end{equation}
since
\begin{equation}
<p| U^{-1}(g') g^a(x) U(g) |p> \neq 0
\end{equation}
for $g' \neq g$.

The lowest order Ward identities -- which are of particular
 relevance for our problem --
\begin{equation}
<p| [g^a({\bf r}, t), g^b({\bf r'}, t)] |p> = 0
\end{equation}
are constraints on 2-, 3- and 4-point Green's functions. Since
$ [g^a({\bf r}, t), g^b({\bf r'}, t)] $ are conserved 4-point operators,
they are also conserved within a 4-point truncation scheme for the Green's
functions. The problem of the Ward identities thus is again shifted to an
initial value problem.

A further question left is related to truncation schemes
in the equations of motion for the field propagators. Here we can directly
refer to the results gained earlier \cite{r5}, i.e. that the conservation
laws for n-point operators are fulfilled if the truncation scheme on
multi-point
Green's functions is performed at order n+1 (assuming all n+1 - point
correlation Green's functions to be zero).
Noticing that $g^a(x)$ contains
1- and 2-gluon-field and 2 quark-field operators, we
can conclude that a correlation
dynamics up to 4-point Green's functions obeys the weak Gauss law.
Thus we have achieved the compatibility of
the gauge constraint (10) with the truncation of the dynamical equations of
motion up to 4-point Green's functions, i.e. the traditional domain of
non-perturbative two-body correlation dynamics. Truncation after the 4-point
functions appears to be natural for non-abelian gauge theories due to the
fact that the bare vertices involve at most four fields (4-gluon
interaction).

Summarizing, we have shown that for a SU(N) gauge theory
in temporal gauge the dynamical evolution of the fields via
Heisenberg equations is compatible with the conservation of the weak Gauss
law and Ward identities up to 4th order in a truncation scheme up to the same
order, provided that the initial conditions follow the gauge constraints.
For example, a functional of the form $F[B^a_i B^a_i]$, where $B^a_i$ denotes
the chromomagnetic field, obeys the Gauss
law \cite{r13}-\cite{r13b} and might be used as a trial wave function in a
variational
calculation within a finite number of basis states. Alternatively, one might
start from perturbative gluon-field configurations and increase the
coupling $g$ from zero adiabatically for the preparation of an initial
state \footnote{Note that the perturbative vacuum $|0>$ is a
 physical state in our
realization of the gauge constraint because it fulfills (10) as well as (27).}.
In this way, one expects to observe a phase transition from the
deconfined to the confined regime if a critical value of $g$ is exceeded.

Inspite of the promising perspectives of the approach discussed here, we
have to point out two problems that remain unsolved so far. First,
the equations for the correlated Green's function are lengthy \cite{Wang93},
 and a
numerical solution might be difficult. However, we expect that this task
is still much less complex than lattice calculations. Secondly, the
correlated Green's functions contain singularities which have to be
renormalized. This may be done partially by normal ordering
the field operators within the Green's functions  and by a mass
renormalization as will be shown in a forthcoming publication \cite{Wang93}.

\vspace{1cm}

One of the authors (S. J. W.) likes to thank Prof. U. Mosel for the kind
hospitality at the University of Giessen where this work was peformed.



\begin{thebibliography}{99}
\bibitem{r1a}{T. Hatsuda, Nucl. Phys. {\bf A544} (1992) 27c.}
\bibitem{r1}{B. Peterson, Nucl. Phys. {\bf A525} (1991) 237c.}
\bibitem{r2}{T.S. Bir\'o, Ann. Phys. (N.Y.) {\bf 191} (1989) 1}
\bibitem{r2a}{M.H. Thoma and H.J. Mang, Z. Phys. {\bf C44} (1989) 349;
             M.H. Thoma, Mod. Phys. Lett. {\bf A7} (1992) 153}
\bibitem{r2b}{J. Ahlbach, M. Lavelle, M. Schaden, and A. Streibl, Phys. Lett.
             {\bf B275} (1992) 124.}
\bibitem{r3a}{H.-Th. Elze, M. Gyulassy and D. Vasak, Phys. Lett. {\bf B177}
             (1986) 402; Nucl. Phys. {\bf B276} (1986) 706.}
\bibitem{r3}{H.-Th. Elze and U. Heinz, Phys. Rep. {\bf 183} (1989) 81.}
\bibitem{r4}{I. Bialynicki-Birula, E.D. Davis, and J. Rafelski, Phys. Lett.
             {\bf B311} (1993) 329.}
\bibitem{r5}{S.J. Wang and W. Cassing, Ann. Phys. (N.Y.) {\bf 159} (1985) 328.}
\bibitem{r6}{S.J. Wang and W. Cassing, Nucl. Phys. {\bf A495} (1989) 371c.}
\bibitem{r7}{W. Cassing and S.J. Wang, Z. Phys. {\bf A337} (1990) 1.}
\bibitem{r8}{S.J. Wang, B.A. Li, W. Bauer, and J. Randrup, Ann. Phys. (N.Y.)
             {\bf 209} (1991) 251.}
\bibitem{r9}{S.J. Wang, W. Zuo, W. Cassing, Giessen University preprint
             (1993).}
\bibitem{r10}{N.H. Christ and T.D. Lee, Phys. Rev. {\bf D22} (1980) 939;
              T.D. Lee, Particle Physics and Introduction to Field Theory,
              Harwood, Chur, 1981.}
\bibitem{r11}{J.M. Cornwall, Phys. Rev. {\bf D38} (1988) 656;
              J.E. Shrauner, J. Phys. {\bf G19} (1993) 979.}
\bibitem{r12}{M.H. Thoma, Ph.D. thesis, TU M\"unchen (1988).}
\bibitem{r13}{J.P. Greensite, Nucl. Phys. {\bf B158} (1979) 469}
\bibitem{r13a}{R.P. Feynman, Nucl. Phys. {\bf B185} (1981) 479}
\bibitem{r13b}{P.E. Haagensen, Barcelona University preprint UB-ECM-PF 93/16
              (1993)}
\bibitem{Wang93}{S. J. Wang, W. Cassing, J. H\"auser, A. Peter and M. H. Thoma,
                 to be published}
\end{thebibliography}
\end{document}